\documentclass[showpacs,amsmath,amssymb,aps,showkeys,floatfix,prd,a4paper]{revtex4}

\usepackage[dvips]{graphicx}
\usepackage{dcolumn}
\usepackage{bm}
\usepackage{epsfig}
\usepackage{amsfonts}
\usepackage{amssymb,amscd}

\def\lsim{\raise0.3ex\hbox{$<$\kern-0.75em\raise-1.1ex\hbox{$\sim$}}}
\def\gsim{\raise0.3ex\hbox{$>$\kern-0.75em\raise-1.1ex\hbox{$\sim$}}}

\def\beq{\begin{equation}}
\def\eeq{\end{equation}}
\def\bea{\begin{eqnarray}}
\def\eea{\end{eqnarray}}
\def\bq{\begin{quote}}
\def\eq{\end{quote}}

\newcommand{\rkp}{\mbox{\boldmath $k_{\perp}$}}

\newcommand{\rs}{\mbox{\boldmath $S$}}

\newcommand{\rp}{\mbox{\boldmath $P$}}

\newcommand{\kp}{\mbox{$k_{\perp}$}}

\newcommand{\rkkp}{\mbox{ $k^2_{\perp}$}}

\newcommand{\rqq}{\mbox{\boldmath $q$}}

\def\gappeq{\mathrel{\rlap {\raise.5ex\hbox{$>$}}
{\lower.5ex\hbox{$\sim$}}}}

\def\lappeq{\mathrel{\rlap{\raise.5ex\hbox{$<$}}
{\lower.5ex\hbox{$\sim$}}}}

\def\Toprel#1\over#2{\mathrel{\mathop{#2}\limits^{#1}}}

\begin{document}


\title{Investigating the Transverse Single Spin Asymmetry in the Inelastic $J/\Psi$ photoproduction in $p^\uparrow p$ and $p^\uparrow A$ collisions}

\author{V.~P. Gon\c{c}alves}
\email{barros@ufpel.edu.br}
\affiliation{High and Medium Energy Group, \\
Instituto de F\'{\i}sica e Matem\'atica, Universidade Federal de Pelotas\\
Caixa Postal 354, CEP 96010-900, Pelotas, RS, Brazil}

\date{\today}

\begin{abstract}
In this paper we propose to investigate the transverse single spin asymmetry in the inelastic $J/\Psi$ photoproduction in $p^\uparrow p$ and $p^\uparrow A$ collisions at RHIC energies. At leading order this process probes the gluon Sivers function. We predict large values for the cross sections, which indicates that its experimental analysis is, in principle,  feasible. The rapidity dependence of the single spin asymmetry is presented. We obtain that the asymmetry is strongly dependent on the model used for the gluon Sivers function and that it can be probed by the analysis of the $J/\Psi$ production at forward rapidities. Our results indicate that a future experimental analysis of this process can be useful to constrain the gluon Sivers function.
\end{abstract}

\maketitle

\section{Introduction}
\label{intro}

The study of high energy processes involving polarized hadrons allows to 
improve our understanding of the polarized quark and gluon structure of the hadrons and the QCD dynamics at a high - energy scale (See e.g. Refs. \cite{eic1,eic2}). In particular, the analysis of the transverse spin phenomena in hard processes is expected to provide a three - dimensional picture of the partons inside the nucleon. One of the current challenges in hadronic physics is the understanding of the large transverse single - spin asymmetries (SSAs), which have been observed in several experiments \cite{her1,her2,comp1,jlab1,jlab2,starexp,comp2,comp3}. A possible explanation for the presence of this asymmetry was proposed many years ago \cite{sivers} and is known as Sivers effect, which considers the correlation between the transverse momentum of partons and the polarization vector of the nucleon. In recent years there has been significant progress in both experimental and theory toward understanding the origin of the SSAs (See e.g. \cite{review_experimental}). In particular, the experimental data released by the HERMES,  COMPASS, Jefferson Lab, PHENIX and STAR Collaborations has allowed the extraction of the Sivers functions for $u$ and $d$ quarks \cite{quark1,quark2,quark3,quark4}.  However, the size of the gluon Sivers function still remains an open question, with no hard constraint existing apart from the positivity bound \cite{review_sivers}.

One process that can be used to probe gluons inside hadrons is the quarkonium production \cite{review_nrqcd}. Several authors have proposed to constrain the gluon Sivers function using the experimental data for the quarkonium production in proton -- proton \cite{sivers_pp} and electron -- proton \cite{godbole,asmita} collisions, considering different initial and final states as e.g. $p^\uparrow p \rightarrow J/\Psi \gamma X$,   $p^\uparrow p \rightarrow J/\Psi J/\Psi X$ and $e p^\uparrow  \rightarrow e^{\prime} J/\Psi \gamma X$ (For a recent review see e.g. Ref. \cite{boer}). These studies indicate that this process is ideal to get a deeper knowledge of the nucleon structure. Our goal in this paper is to complement these previous analysis and propose the study of the gluon Sivers function in the photoproduction of vector mesons in $p^\uparrow p$ and  $p^\uparrow A$ collisions at high energies.  During the last years, the study of  photon -- induced interactions  \cite{upc} at Tevatron, RHIC and LHC  became a reality \cite{cdf,star,phenix,alice,alice2,lhcb,lhcb2,lhcb3,lhcbconf} and new data are expected to be released soon (For a recent review see e.g. \cite{review_forward}). One the main motivations to the study of these processes is the possibility to constrain the main aspects of the treatment of the QCD dynamics at high energies and large nuclei (See e.g. Refs. \cite{bert,vicmag,brunoall,brunorun2,vicdiego,roman}). These previous analysis have been performed considering the collision of unpolarized hadrons. Here we extend these studies for the case where one transversely polarized proton beam is present and estimate the impact of different models for the gluon Sivers function on the transverse single spin asymmetry.

The basic idea in photon -- induced interactions is that 
 an ultra relativistic charged hadron (proton or nucleus) 
gives rise to strong electromagnetic fields, such that the photon stemming from the electromagnetic field of one of the two colliding hadrons can 
interact with one photon of the other hadron (photon - photon process) or can interact directly with the other hadron (photon - hadron process) \cite{upc}. 
In these processes the total cross section  can be factorized in terms of the equivalent flux of photons into the hadron projectile and the photon-photon 
or photon-target  cross section.  In the particular case of the inelastic $J/\Psi$ photoproduction in $p^\uparrow p$ and $p^\uparrow A$ collisions, we will assume that the unpolarized hadron ($p$ or $A$) is the source of photons, which interact with the transversely polarized protons at high energies, producing a $J/\Psi$ and dissociating the proton target. In the nuclear case, such approximation is justified due to   enhancement by a factor $Z^2$  in the nuclear photon flux in comparison to that for a proton (see below), which implies that the photon - induced interactions are dominated by photons from the nucleus. In the case of $p^\uparrow p$ collisions, the process of interest can be separated by tagging the unpolarized proton in the final state, which is present when it emits the photon. As a consequence, the hadronic cross section will be factorized as follows
\begin{eqnarray}
\sigma_{h p^\uparrow  \rightarrow h J/\Psi X} (\sqrt{s})  = \int d x_\gamma d^2\rkp_\gamma \,\,  f_{\gamma/h} (x_\gamma,\rkp_\gamma) \cdot \sigma_{\gamma p^\uparrow  \rightarrow  J/\Psi X} (W_{\gamma p}^2) \,\,,
\label{sig} 
\end{eqnarray}
where $x_\gamma$ is the energy fraction of hadron carried by the photon with transverse momentum $\rkp_\gamma$ and $f_{\gamma/h}$ is the photon flux associated to hadron $h$. Moreover,   $W_{\gamma h}$ is the c.m.s. photon-proton energy given by $W_{\gamma p}=[2\,\omega\sqrt{s}]^{1/2}$, where $\omega$ is the photon energy and $\sqrt{s}$ is  the c.m.s energy of the
hadron-proton system. The final state will be characterized by the presence of one rapidity gap and an intact hadron, which we assume to be the unpolarized one. Both aspects can be used in principle to experimentally separate the vector mesons produced by photon -- induced interactions. In our  
exploratory study we will assume that the transverse momentum dependence of the photon distribution can be described by a simple Gaussian form: 
$f_{\gamma/h} (x_\gamma,\rkp_\gamma) = f_{\gamma/h} (x_\gamma) \exp{(-\rkkp_\gamma/\langle\rkkp_\gamma\rangle}) / (\pi \langle\rkkp_\gamma\rangle)$. Moreover, we will assume that 
the  photon spectrum $f_{\gamma/h} (x_\gamma)$ associated to a proton is given by  \cite{Dress},
\begin{eqnarray}
f_{\gamma/p} (x_\gamma) =  \frac{\alpha_{\mathrm{em}}}{2 \pi}  
\frac{1 + (1-x_\gamma)^2}{x_\gamma}
\left( \ln{\Omega} - \frac{11}{6} + \frac{3}{\Omega}  - \frac{3}{2 \,\Omega^2} + \frac{1}{3 \,\Omega^3} \right) \,,
\label{eq:photon_spectrum}
\end{eqnarray}
with the notation $\Omega = 1 + [\,(0.71 \,\mathrm{GeV}^2)/Q_{\mathrm{min}}^2\,]$ and $Q_{\mathrm{min}}^2= m_p^2  x_\gamma^2/(1-x_\gamma)$. This expression  is derived considering the Weizs\"{a}cker-Williams method of virtual photons and using an elastic proton form factor (For more details see Refs. \cite{Dress,Kniehl}).
In the case of $pA$ collisions, 
an analytic approximation for the equivalent photon flux of a nuclei can be calculated considering the requirement that  photoproduction
is not accompanied by hadronic interaction (ultra-peripheral
collision), which is given by \cite{upc}
\begin{eqnarray}
f_{\gamma/A} (x_\gamma) = \frac{\alpha_{em} Z^2}{\pi}\, \frac{1}{x_\gamma} \left[2 {\eta}\,K_0\,({\eta})\, K_1\,({\eta}) - {\eta}^2\,{\cal{U}}({\eta}) \right]\,
\label{fluxint}
\end{eqnarray}
where   $K_0(\eta)$ and  $K_1(\eta)$ are the
modified Bessel functions, ${\eta}= x_\gamma \, m_p \, b_{min}$ and  ${\cal{U}}({\eta}) = K_1^2\,({\eta})-  K_0^2\,({\eta})$. In our analysis we will assume that 
$b_{min} = R_{p}+R_{A}$, which suppress the strong interactions.

In order to estimate the inelastic $J\Psi$ photoproduction in hadronic collisions, we should to specify the underlying mechanism governing heavy quarkonium production, which is still a subject of intense debate. As reviewed in Ref. \cite{review_nrqcd},  a number of theoretical approaches have been proposed in the last years for the calculation of the heavy quarkonium production, as for instance,  the Non Relativistic QCD (NRQCD)  approach, the fragmentation approach, the color singlet model (CSM), the Color Evaporation Model and the $k_T$-factorization approach. In our analysis we will assume the Color Evaporation Model \cite{velhos_cem,cem_gregores,criscem,crisvic_cem,ramona}, generalized to take into account the transverse momentum dependence of the gluon distribution function \cite{godbole}. 
 Our motivation to use this model is associated to its simplicity and to the fact that we have checked that this model is able to describe the HERA data for the inelastic $J/\Psi$ photoproduction (See also Ref. \cite{cem_gregores}). The basic idea in the Color Evaporation Model is that
the formation of the color singlet state is not enforced at the perturbative level. The cross section for the process is given essentially by the boson-gluon cross section and
the assumption that the color neutralization of the $Q \bar{Q}$ occurs by interaction with the surrounding color field. 
In CEM, quarkonium production is treated identically to open
heavy quark production with exception that in the case of
quarkonium, the invariant mass of the heavy quark pair is
restricted to be below the open meson threshold, which is twice
the mass of the lowest meson mass  that can be formed with the
heavy quark. Moreover, the CEM assumes
that the quarkonium dynamics is identical to all quarkonium
states, although the $Q\overline{Q}$ pairs are typically produced
at short distances in different color, angular momentum and spin
states. In the case of charmonium states, the hadronization from the
$c\overline{c}$ pairs is nonperturbative, usually involving the
emission of one or more soft gluons. Depending on the quantum
numbers of the initial $c\overline{c}$ pair and the final state, a different matrix element is needed for the
production of the charmonium state. The average of these
nonperturbative matrix elements are combined into the universal
factor $F[nJ^{PC}]$, which is process- and kinematics-independent
and describes the probability that the $c\overline{c}$ pair binds
to form a quarkonium of a given spin $J$,
parity $P$, and  charge conjugation $C$.
Considering the $J/\Psi$ production in  $\gamma p^\uparrow $ interactions, the CEM predicts that the cross section will be given by 
\begin{eqnarray}
\sigma_{\gamma p^\uparrow \rightarrow J/\Psi X} = F_{J/\Psi} \,\,\overline{\sigma}_{\gamma p^\uparrow \rightarrow c
\overline{c} X} \,\,,
\label{sigcem}
\end{eqnarray}
where the short distance contribution is
\begin{eqnarray}
\overline{\sigma}_{\gamma p^\uparrow \rightarrow c
\overline{c} X} = \int_{4m_c^2}^{4 m_D^2}
dM^2_{c\overline{c}} \, dx_g \, d^2 \rkp_g \, f_{g/p^\uparrow} (x_g, \rkp_g) 
\,\,\frac{d\sigma[\gamma g \rightarrow c\overline{c}]}{dM^2_{c\overline{c}}} \,\,,
\label{cross}
\end{eqnarray}
where  $M_{c\overline{c}}$
is the invariant mass of
the $c\overline{c}$ pair, $m_c$ is the charm quark mass and $2m_D$ is the $D%
\overline{D}$ threshold. One have that the cross section for the inelastic $J/\Psi$ photoproduction is  proportional to the number density of gluons inside a proton with transverse polarization $\rs$ and momentum $\rp$, which is usually parameterized as \cite{anselmino} 
\begin{eqnarray}
f_{g/p^\uparrow} (x_g, \rkp_g,\rs) \equiv f_{g/p} (x_g, \kp_g) + \frac{1}{2} \Delta^N f_{g/p^\uparrow} (x_g, \kp_g) \hat{\rs} \cdot (\hat{\rp} \times \hat{\rkp}_g) \,\,,
\label{fgeral} 
\end{eqnarray}
where $x_g$ is the longitudinal momentum fraction of the gluon and $\rkp_g$ its transverse momentum. Moreover, $f_{g/p} (x_g, \kp_g)$ is the unpolarized transverse momentum dependent (TMD)  gluon distribution and $  
\Delta^N f_{g/p^\uparrow} (x_g, \kp_g)$ is the gluon Sivers function. 

In order to probe the gluon Sivers function in the inelastic $J/\Psi$ photoproduction in
$p^\uparrow p$ and  $p^\uparrow A$ collisions, in what follows we will investigate the impact of different models for $  
\Delta^N f_{g/p^\uparrow} (x_g, \kp_g)$ in the rapidity ($Y$) dependence of the single spin asymmetry, defined as
\begin{eqnarray}
A_N (Y) = \frac{\frac{d\sigma^\uparrow}{dY} - \frac{d\sigma^\downarrow}{dY}}{\frac{d\sigma^\uparrow}{dY} + \frac{d\sigma^\downarrow}{dY}} \,\,,
\label{any}
\end{eqnarray} 
where $\frac{d\sigma^\uparrow}{dY}$ and $\frac{d\sigma^\downarrow}{dY}$ are respectively
the differential cross sections measured when the proton is transversely polarized up ($\uparrow$) and down ($\downarrow$) with respect to the scattering plane, calculated using Eqs. (\ref{sig}), (\ref{sigcem}) and (\ref{fgeral}). 
 As in Ref. \cite{godbole} we will estimate the numerator with a weight factor $\sin (\phi_{q_T} - \phi_S)$, where $\phi_{q_T}$ and $\phi_S$ are the azimuthal angles of the $J/\Psi$ and proton spin, respectively.  One have that \cite{godbole}
 \begin{eqnarray}
 \frac{d\sigma^\uparrow}{dY} - \frac{d\sigma^\downarrow}{dY} & = & F_{J/\Psi} \, \int d\phi_{q_T} \int q_T dq_T \int_{4m_c^2}^{4 m_D^2}
dM^2_{c\overline{c}} \int d^2 \rkp_g \, f_{\gamma/h} (x_\gamma, \rqq_T - \rkp_g)  \nonumber \\
& \times & [f_{g/p^\uparrow}(x_g, \rkp_g)  - f_{g/p^\downarrow}(x_g, \rkp_g)] \, 
\hat{\sigma}_0 (M^2_{c\overline{c}}) \,\sin (\phi_{q_T} - \phi_S)
\label{num}
 \end{eqnarray}
 and 
 \begin{eqnarray}
 \frac{d\sigma^\uparrow}{dY} + \frac{d\sigma^\downarrow}{dY} = 2 \, F_{J/\Psi} \, \int d\phi_{q_T} \int q_T dq_T \int_{4m_c^2}^{4 m_D^2}
dM^2_{c\overline{c}} \int d^2 \rkp_g \, f_{\gamma/h} (x_\gamma, \rqq_T - \rkp_g) \,   f_{g/p} (x_g, \rkp_g) \hat{\sigma}_0(M^2_{c\overline{c}}) \,\,, 
\label{den}
 \end{eqnarray}
 where $q_T$ is the transverse momentum of the vector meson and $\hat{\sigma}_0$ is the partonic cross section for the $\gamma g \rightarrow c \bar{c}$ process \cite{gluck_reya}. It is important to emphasize that the spin asymmetry is not dependent on $F_{J/\Psi}$. 
 Our motivation to investigate the rapidity dependence of $A_N$ is associated to the fact that the rapidity $Y$ of the vector meson determines the typical values of $x_\gamma$ and $x_g$ probed in the interaction, which are given by $x_{g,\gamma} = M_{c\overline{c}}/\sqrt{s} \exp(\pm Y)$. Therefore, its analysis allow us to know the value of $x_g$ that is being probed in the gluon Sivers function. 
  In what follows we will assume that the unpolarized TMD gluon distribution $f_{g/p} (x_g, \rkp_g)$ can be described by a Gaussian form:
 \begin{eqnarray}
 f_{g/p} (x_g, \rkp_g) = f_{g/p} (x_g, \mu^2) \frac{1}{\pi \langle \rkkp_g \rangle} e^{-\rkkp_g / \langle \rkkp_g \rangle}
 \end{eqnarray}
 with the factorization scale $\mu^2$ being given by $M^2_{c\overline{c}}$. As in Refs. \cite{pisano,godbole}, we choose a frame where the proton is moving along the $z$ -- axis  with momentum $\rp$, is transversely polarized along $y$ -- axis and $\rkp = k_\perp (\cos \phi_{k_{\perp,g}}, \sin \phi_{k_{\perp,g}}, 0)$, which implies that  
 $\hat{\rs} \cdot (\hat{\rp} \times \hat{\rkp}_g) = \cos \phi_{k_{\perp,g}}$.
 Moreover, we will consider that the gluon Sivers function can be described as follows
 \begin{eqnarray}
  \Delta^N f_{g/p^\uparrow} (x_g, \kp_g) = 2 N_g(x_g) f_{g/p} (x_g,\mu^2) h(\kp_g)  \frac{e^{-\rkkp_g / \langle \rkkp_g \rangle}}{\pi \langle \rkkp_g \rangle} \,\,, 
  \label{sivers_pisano}
 \end{eqnarray}
 where 
 \begin{eqnarray}
 N_g(x_g) = N_g x_g^\alpha (1-x_g)^\beta \frac{(\alpha + \beta)^{(\alpha + \beta)}}{\alpha^\alpha \beta^\beta}
 \end{eqnarray}
 with $|N_g| \le 1$ and 
 \begin{eqnarray}
 h(\kp_g) = \sqrt{2 e} \, \frac{\kp_g}{M^\prime} e^{-\rkkp_g/M^{\prime 2}}\,\,.
 \end{eqnarray}
 The $k_\perp$ dependent part of the Sivers function can expressed as follows
 \begin{eqnarray}
    h(\kp_g) \, \frac{e^{-\rkkp_g / \langle \rkkp_g \rangle}}{\pi \langle \rkkp_g \rangle} = \frac{\sqrt{2 e}}{\pi} \sqrt{\frac{1 - \rho}{\rho}} \, \kp_g \frac{e^{-\rkkp_g/\rho \langle \rkkp_g \rangle }}{\langle \rkkp_g \rangle^{3/2}} \,\,,
\end{eqnarray}  
where $\rho \equiv M^{\prime 2} / ( \langle \rkkp_g \rangle + M^{\prime 2})$.  
The parametrization given by Eq. (\ref{sivers_pisano}) was proposed in Ref. \cite{quark1} and recently used in Ref. \cite{pisano},  where the authors have considered the midrapidity data on the transverse single spin asymmetry measured in $pp \rightarrow \pi^0 X$ by the PHENIX Collaboration at RHIC  and the present information on the quark Sivers functions to get a first estimate on the gluon Sivers distribution. Assuming $\langle \rkkp_g \rangle = 0.25$ GeV$^2$, they have obtained two different sets for the best -- fit parameters $N_g$, $\alpha$, $\beta$ and $\rho$,  denoted by SIDIS1 and SIDIS2 (For details see \cite{pisano}).

\begin{figure}[t]
\begin{tabular}{ccc}
\includegraphics[scale=0.35]{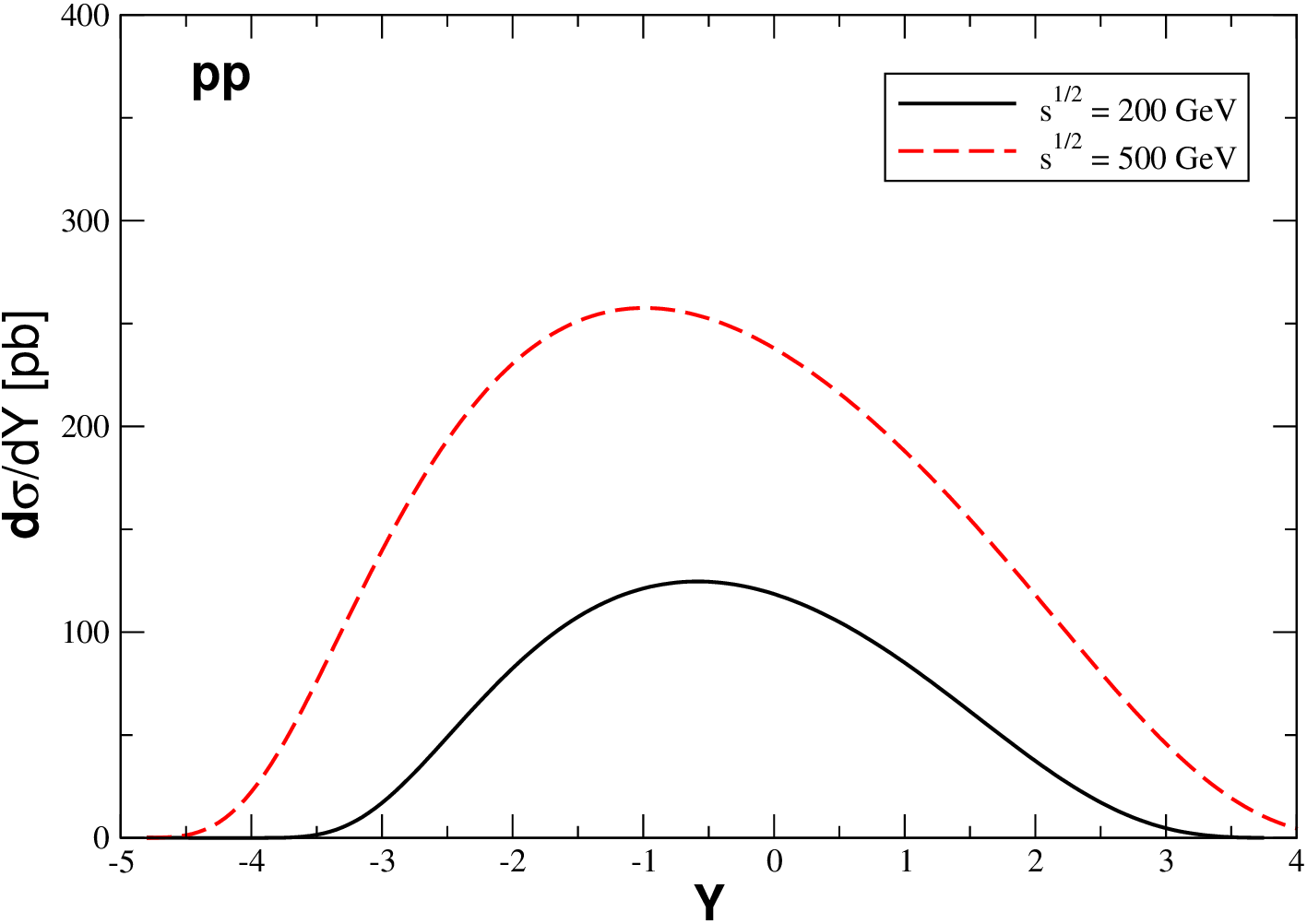} & \,\,\,&
\includegraphics[scale=0.35]{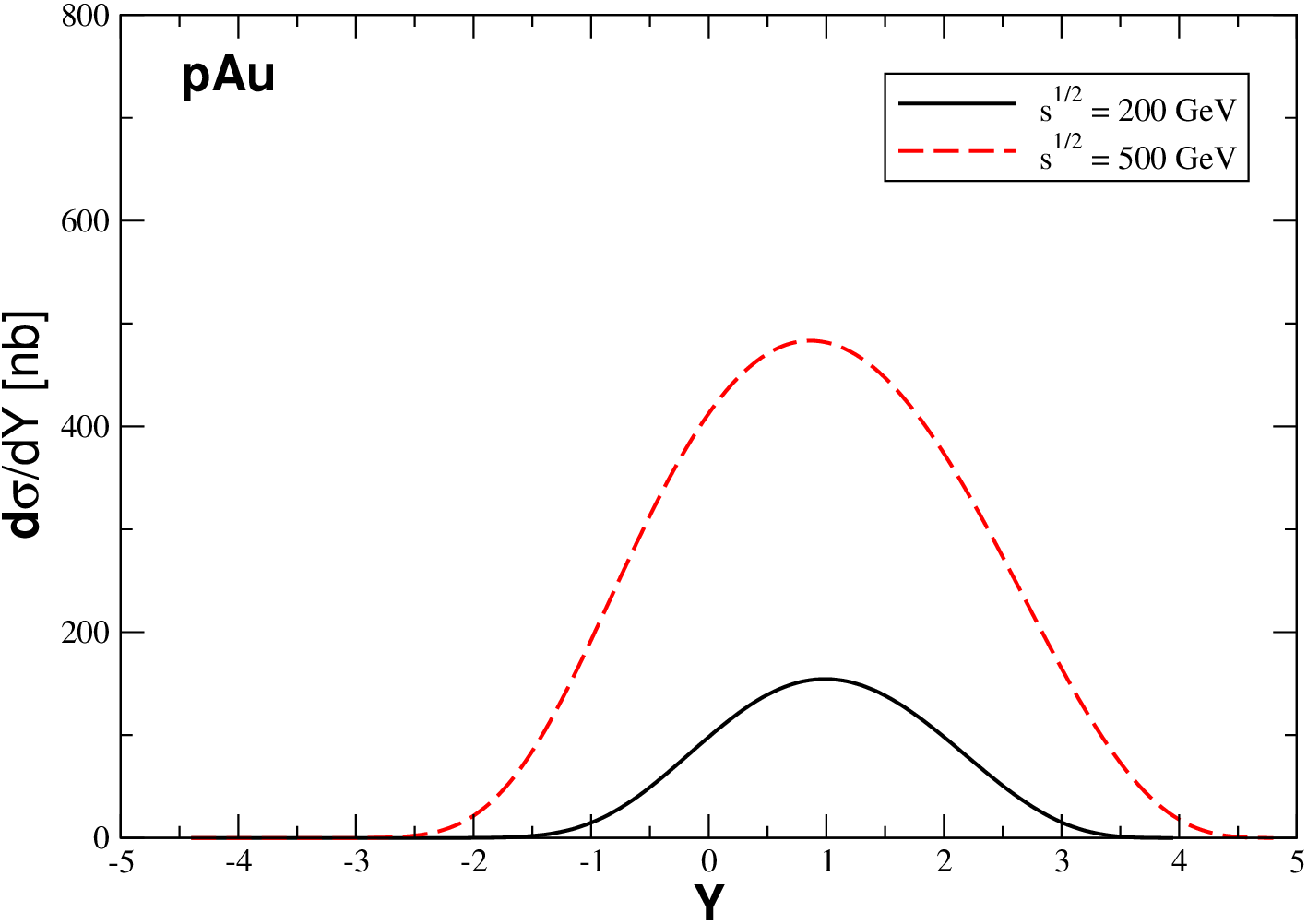} 
\end{tabular}
\caption{Rapidity distribution for the inelastic $J/\Psi$ photoproduction in $pp$ (left panel) and $pAu$ (right panel) collisions  at $\sqrt{s} = 200$  and 500 GeV.}
\label{fig:1}
\end{figure}

\begin{table}[t]
\centering
\begin{tabular}{|c|c|c|}\hline

                              &   $\sqrt{s} = 200$ GeV  & $\sqrt{s} = 500$ GeV                   \\ \hline
$p^\uparrow p$    &    0.932            &   1.245                      \\ \hline
$p^\uparrow Au$   &    380.0             & 1664.5     \\ \hline

\end{tabular} 

\caption{Total cross sections  for the inelastic $J/\Psi$ photoproduction in $pp/pAu$ collisions at RHIC energies. Values in nb.}  
\label{tab1}
\end{table}

In what follows we will present our predictions for $A_N(Y)$ considering the inelastic $J/\Psi$ photoproduction in $p^\uparrow p$ and $p^\uparrow Au$ collisions at different values of the center -- of -- mass energy. We will consider the SIDIS1 and SIDIS2 models for the gluon Sivers function. In order to estimate the impact of different  gluon Sivers distributions on $A_N(Y)$, we also will consider two alternative models obtained assuming that \cite{boervolg} (a) $N_g(x) = [N_u(x) + N_d(x)]/2$ and (b) $N_g(x) = N_u(x)$,   which we will denote by BV-a and BV-b hereafter. In our study we will consider the best fit parameters for the $u$ and $d$ quark Sivers functions obtained recently in Ref. \cite{quark3} from the latest SIDIS data. Moreover, we will integrate the transverse momentum of the vector meson in the range $0 \le q_T \le 1.0$ GeV, assume that $m_c = 1.5$ GeV and $m_D = 1.864$ GeV and use the CTEQ6LO parametrization \cite{cteq6} for the unpolarized gluon distribution ($f_{g/p}$). Although the predictions for the rapidity distributions and total cross sections are sensitive to the modelling of $f_{g/p}$ and $m_c$ (See e.g. \cite{vicmairon}), we have verified that the predictions for the rapidity distributions our results for $A_N (Y)$ are almost independent of these choices. 

Initially lets present in Fig. \ref{fig:1}, by the first time, our predictions for the rapidity distribution considering $pp$ and $pAu$ collisions at $\sqrt{s} = 200$ and 500 GeV. We will assume that $F_{J/\Psi} = (1/9) \cdot \rho_{J/\Psi}$, where the factor $1/9$ represent the statistical probability that the $c\bar{c}$ will be in a color singlet state asymptotically and $\rho_{J/\Psi}$ is a non - perturbative parameter, determined by fitting the data. As in Refs. \cite{criscem, crisvic_cem} we will assume   that $\rho_{J/\Psi} = 0.5$.  In the case of $pp$ collisions, the rapidity distribution shown have been obtained assuming that one of the incident protons acts as the photon source and the other as target. Such assumption implies an asymmetric distribution for a symmetric collision. On the other hand, for $pAu$ collisions, the distribution is asymmetric due to $Z^2$ enhancement present in the nuclear photon flux. The predictions for the total cross sections are presented in Table \ref{tab1}. As expected, the total cross sections for $pAu$ collisions are larger than for $pp$ one.  In comparison to the predictions for the exclusive vector meson production presented e.g. in Ref. \cite{vicmag}, the inelastic production is smaller by a factor $\ge 4$, in agreement with the results obtained in Ref. \cite{vicmairon}. However,  the final state for inelastic production is distinct of the exclusive case and, as indicated in Ref. \cite{brunoluizvictor}, the analysis of the transverse momentum distribution of the vector meson are different, the separation of the inclusive and exclusive contributions is, in principle, feasible. Finally, we have verified that the predictions for the rapidity distributions and total cross sections are modified by $\approx 15 \%$ if the NRQCD formalism is used to describe the quarkonium production, in agreement with the results presented in Ref. \cite{brunoluizvictor}.


\begin{figure}[t]
\begin{tabular}{ccc}
\includegraphics[scale=0.35]{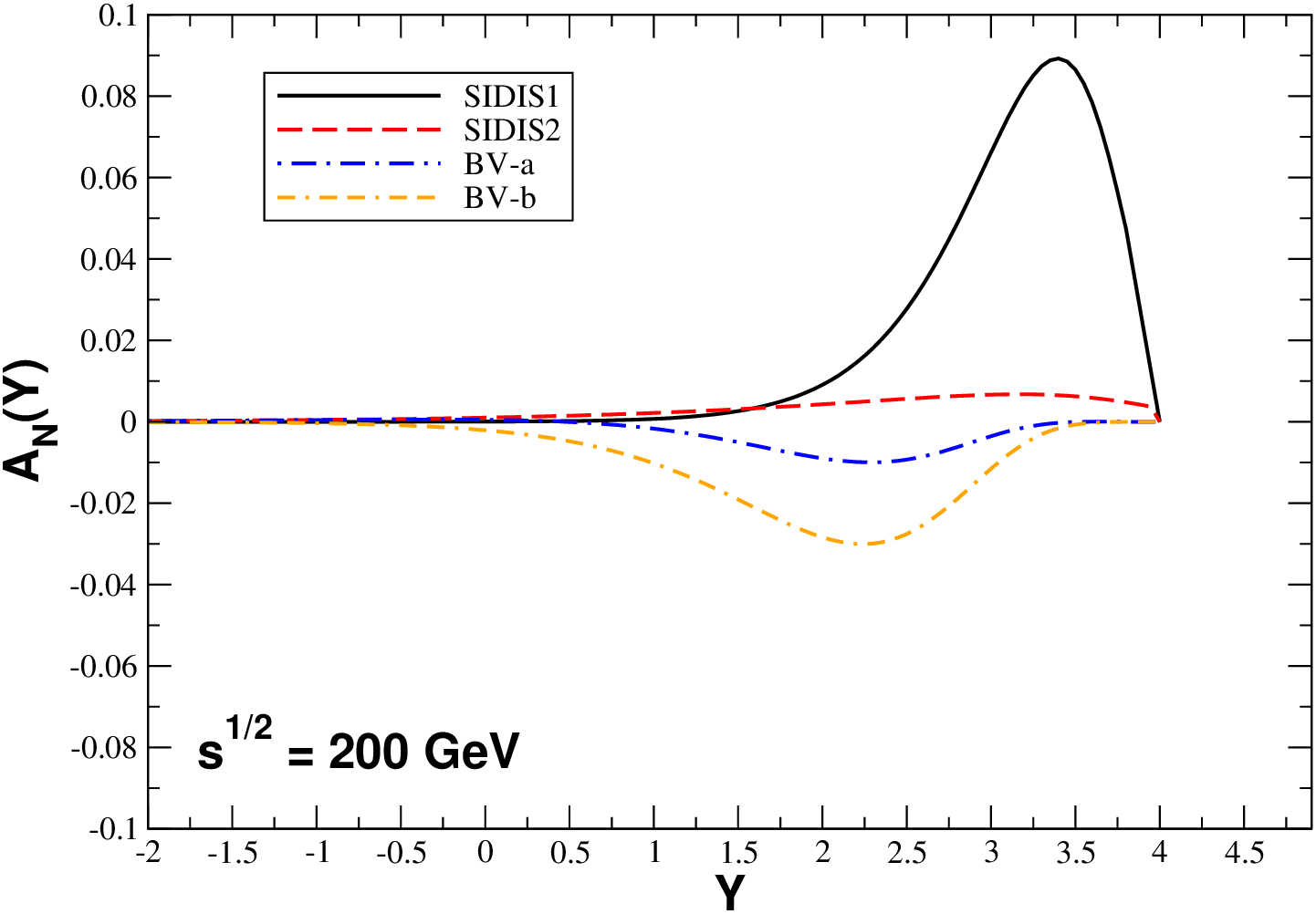} & \,\,\,&
\includegraphics[scale=0.35]{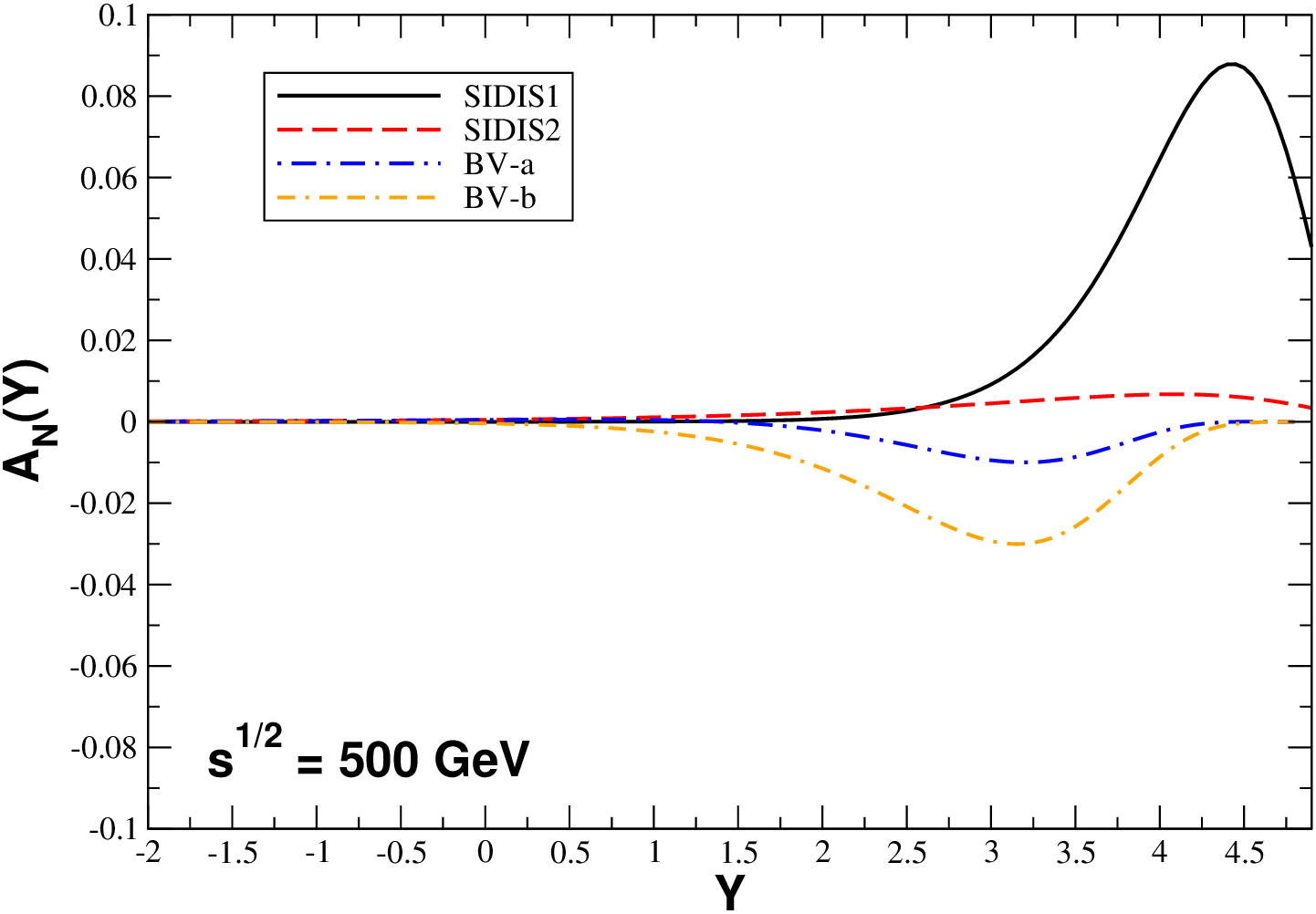} 
\end{tabular}
\caption{Predictions for the single spin asymmetry in the inelastic $J/\Psi$ photoproduction in $p^\uparrow p$ collisions at $\sqrt{s} = 200$ (left panel) and 500 GeV (right panel) considering  different models for the gluon Sivers function.}
\label{fig:2}
\end{figure}

Our predictions for the single spin asymmetry are presented in Fig. \ref{fig:2} considering $p^\uparrow p$ collisions at $\sqrt{s} = 200$ GeV (left panel) and 500 GeV (right panel). We have that the magnitude and signal of $A_N(Y)$ is strongly dependent on the model used for the gluon Sivers function, with the position of the peak ocurring at larger values of $Y$ with the increasing of the energy. Moreover, we have that the maximum and minimum values of $A_N$ are almost independent of energy. These results are consistent with those obtained in Refs. \cite{godbole} for the $J/\Psi$ production in $ep^\uparrow$ collisions.   Our results indicate that the signal and magnitude of the asymmetry can be probed by the analysis of the $J/\Psi$ production at forward rapidities. Additionally, we also have estimated $A_N$ for $p^\uparrow Au$ collisions and obtained that its rapidity dependence, position of the peak and value of  the maximum and minimum are very similar to those obtained in 
$p^\uparrow p$ collisions. Such behaviour is expected, since the photon flux is present in the numerator and denominator of Eq. (\ref{any}), which implies that the $Z^2$ enhancement of the nuclear flux does not affect $A_N$. Therefore,  we predict similar asymmetry in $p^\uparrow p$ and $p^\uparrow Au$  collisions. However, it is important to emphasize that the magnitude of the rapidity distribution in nuclear collisions is almost three orders of magnitude larger than in proton - proton collisions (See Fig. \ref{fig:1}), which implies that the study of  the single spin asymmetry in   $p^\uparrow Au$ collisions is  expected to be more easily  performed. 

Some comments are in order. In our exploratory study we have considered the Color Evaporation Model to describe the quarkonium production.  As pointed before, this subject is still a theme of intense debate. We have verified that if the quarkonium production is treated using the NRQCD formalism, the difference in our predictions for $A_N$ is smaller than $5 \%$. Such small difference in $A_N$ is expected since we are estimating a ratio between cross sections. Similar results have been obtained in  Refs. \cite{godbole,asmita} for $J/\Psi$ production in $ep^\uparrow$ collisions. One aspect that deserves more detailed studies is the analyis is the inclusion of the QCD evolution in the TMD gluon distribution (See e.g. \cite{godbole,asmita}). We postpone the analysis of this topic for a future study. Additionally,   we would like to emphasize that the analysis of the  inelastic $J/\Psi$ photoproduction in $ p^\uparrow p / p^\uparrow A$ collisions should also be possible in the AFTER@LHC experiment \cite{after}. Considering the planned characteristics of the experiment (high luminosity, fixed -- target collisions, ...), and that our results indicate that the maximum value of $A_N$ is almost energy independent, with the peak occuring at large rapidities, we believe that the study of this process is feasible in this experiment. We plan to present more detailed results in a future publication.
Finally, it is important to emphasize that in our analysis we only have considered the central values of the parameters obtained in Refs. \cite{quark3,pisano}. As already pointed out in \cite{pisano} and carefully estimated in Ref. \cite{quark3}, the current uncertainty in these parameters still is large, which has direct impact on  the modelling of the gluon Sivers function. As a consequence, the values for $A_N$ obtained in our analysis and presented in Fig. \ref{fig:2} should be considered illustrative of the potential of the inelastic $J/\Psi$ photoproduction in $ p^\uparrow p / p^\uparrow A$ collisions as a probe of gluon Sivers function. Our hope is that the results presented here motivate a future experimental analysis.




Finally, lets summarize our main results and conclusions. During the last years, the experimental results from Tevatron, RHIC and LHC have demonstrated that the study of  hadronic physics using photon induced interactions in $pp/pA/AA$ colliders is feasible.  In this paper we have estimated, by the first time, the inelastic $J/\Psi$ photoproduction in $ p^\uparrow p / p^\uparrow A$ collisions at RHIC energies. Moreover, the impact of different models for the gluon Sivers function on the transverse single spin asymmetry have been investigated. Our results indicate that the asymmetry is strongly dependent on the modelling of the gluon Sivers function. Moreover, the signal and magnitude of the asymmetry can by investigated by the analysis of the $J/\Psi$ production at forward rapidities. Such aspects motivate a future experimental analysis of this process as a probe of the gluon Sivers function.


\section*{Acknowledgements}
The author is grateful to the members of the THEP group for the hospitality at  Lund University, where the revised version of this work was finished.  This work was partially financed by the Brazilian funding agencies  CNPq,  FAPERGS and INCT-FNA (process number 464898/2014-5).


\begin{thebibliography}{99}

\bibitem{eic1} 
  D.~Boer, M.~Diehl, R.~Milner, R.~Venugopalan, W.~Vogelsang, D.~Kaplan, H.~Montgomery and S.~Vigdor {\it et al.},
  arXiv:1108.1713 [nucl-th].
  
  \bibitem{eic2} 
  A.~Accardi, J.~L.~Albacete, M.~Anselmino, N.~Armesto, E.~C.~Aschenauer, A.~Bacchetta, D.~Boer and W.~Brooks {\it et al.},
 Eur.\ Phys.\ J.\ A {\bf 52}, no. 9, 268 (2016)
 
\bibitem{her1} 
  A.~Airapetian {\it et al.} [HERMES Collaboration],
  Phys.\ Rev.\ Lett.\  {\bf 94}, 012002 (2005) 
 
 \bibitem{her2} 
  A.~Airapetian {\it et al.} [HERMES Collaboration],
  Phys.\ Rev.\ Lett.\  {\bf 103}, 152002 (2009)
 
 \bibitem{comp1} 
  C.~Adolph {\it et al.} [COMPASS Collaboration],
  Phys.\ Lett.\ B {\bf 717}, 383 (2012)
 
 \bibitem{jlab1} 
  X.~Qian {\it et al.} [Jefferson Lab Hall A Collaboration],
  Phys.\ Rev.\ Lett.\  {\bf 107}, 072003 (2011)
 
 
 
 \bibitem{jlab2} 
  Y.~X.~Zhao {\it et al.} [Jefferson Lab Hall A Collaboration],
  Phys.\ Rev.\ C {\bf 90}, no. 5, 055201 (2014)
 
\bibitem{starexp} 
  L.~Adamczyk {\it et al.} [STAR Collaboration],
  Phys.\ Rev.\ Lett.\  {\bf 116}, no. 13, 132301 (2016) 
 
 \bibitem{comp2} 
  C.~Adolph {\it et al.} [COMPASS Collaboration],
  Phys.\ Lett.\ B {\bf 772}, 854 (2017)
 
 \bibitem{comp3} 
  C.~Adolph {\it et al.} [COMPASS Collaboration],
  Phys.\ Lett.\ B {\bf 770}, 138 (2017)
 
 
 
 
\bibitem{sivers} 
  D.~W.~Sivers,
  Phys.\ Rev.\ D {\bf 41}, 83 (1990).


 \bibitem{review_experimental} 
  E.~C.~Aschenauer, U.~D'Alesio and F.~Murgia,
  Eur.\ Phys.\ J.\ A {\bf 52}, no. 6, 156 (2016)


\bibitem{quark1} 
  M.~Anselmino, M.~Boglione, U.~D'Alesio, A.~Kotzinian, F.~Murgia and A.~Prokudin,
  Phys.\ Rev.\ D {\bf 72}, 094007 (2005)
  Erratum: [Phys.\ Rev.\ D {\bf 72}, 099903 (2005)]

\bibitem{quark2} 
  M.~Anselmino, V.~Barone and M.~Boglione,
  Phys.\ Lett.\ B {\bf 770}, 302 (2017)

\bibitem{quark3} 
  M.~Anselmino, M.~Boglione, U.~D'Alesio, F.~Murgia and A.~Prokudin,
  JHEP {\bf 1704}, 046 (2017)

\bibitem{quark4} 
  A.~Martin, F.~Bradamante and V.~Barone,
  Phys.\ Rev.\ D {\bf 95}, no. 9, 094024 (2017)

\bibitem{review_sivers} 
  D.~Boer, C.~Lorcé, C.~Pisano and J.~Zhou,
  Adv.\ High Energy Phys.\  {\bf 2015}, 371396 (2015)



\bibitem{review_nrqcd}
  N.~Brambilla, S.~Eidelman, B.~K.~Heltsley, R.~Vogt, G.~T.~Bodwin, E.~Eichten, A.~D.~Frawley and A.~B.~Meyer {\it et al.},
  Eur.\ Phys.\ J.\ C {\bf 71}, 1534 (2011)


\bibitem{sivers_pp}
M.~Anselmino, M.~Boglione, U.~D'Alesio, E.~Leader and F.~Murgia,
  Phys.\ Rev.\ D {\bf 70}, 074025 (2004);  M.~Anselmino, V.~Barone, A.~Drago and N.~N.~Nikolaev,
  Phys.\ Lett.\ B {\bf 594}, 97 (2004); 
D.~Boer and C.~Pisano,
  Phys.\ Rev.\ D {\bf 86}, 094007 (2012);  G.~P.~Zhang,
  Phys.\ Rev.\ D {\bf 90}, no. 9, 094011 (2014); A.~Mukherjee and S.~Rajesh,
  Phys.\ Rev.\ D {\bf 93}, no. 5, 054018 (2016);U.~D'Alesio, F.~Murgia, C.~Pisano and P.~Taels,
  Phys.\ Rev.\ D {\bf 96}, no. 3, 036011 (2017);  R.~M.~Godbole, A.~Kaushik, A.~Misra, V.~Rawoot and B.~Sonawane,
  arXiv:1703.01991 [hep-ph].
  
\bibitem{godbole} 
  R.~M.~Godbole, A.~Misra, A.~Mukherjee and V.~S.~Rawoot,
  Phys.\ Rev.\ D {\bf 85}, 094013 (2012); Phys.\ Rev.\ D {\bf 88}, no. 1, 014029 (2013); R.~M.~Godbole, A.~Kaushik, A.~Misra and V.~S.~Rawoot,
  Phys.\ Rev.\ D {\bf 91}, no. 1, 014005 (2015)
  
\bibitem{asmita}
A.~Mukherjee and S.~Rajesh,
  arXiv:1609.05596v1 [hep-ph].    
  



\bibitem{boer} 
  D.~Boer,
  Few Body Syst.\  {\bf 58}, no. 2, 32 (2017)





\bibitem{upc}
 G. Baur, K. Hencken, D. Trautmann, S. Sadovsky, Y. Kharlov, Phys.
Rep. {\bf 364}, 359 (2002); 
V.~P.~Goncalves and M.~V.~T.~Machado,
Mod. Phys. Lett. A {\bf 19}, 2525  (2004); 
 C.~A. Bertulani, S.~R.~Klein and J.~Nystrand, Ann. Rev. Nucl. Part. Sci. {\bf 55}, 
271 (2005);
 K.~Hencken {\it et al.},
  Phys.\ Rept.\  {\bf 458}, 1 (2008). 



\bibitem{cdf} 
  T.~Aaltonen {\it et al.}  [CDF Collaboration],
  Phys.\ Rev.\ Lett.\  {\bf 102}, 242001 (2009)
  
\bibitem{star} 
  C.~Adler {\it et al.}  [STAR Collaboration],
  Phys.\ Rev.\ Lett.\  {\bf 89}, 272302 (2002)
  
  \bibitem{phenix} 
  S.~Afanasiev {\it et al.}  [PHENIX Collaboration],
  Phys.\ Lett.\ B {\bf 679}, 321 (2009)

\bibitem{alice} 
  B.~Abelev {\it et al.}  [ALICE Collaboration],
  Phys.\ Lett.\ B {\bf 718}, 1273 (2013)


\bibitem{alice2} 
  E.~Abbas {\it et al.}  [ALICE Collaboration],
  Eur.\ Phys.\ J.\ C {\bf 73}, 2617 (2013)
  
\bibitem{lhcb} 
  R. Aaij {\it et al.}  [LHCb Collaboration],
  J.\ Phys.\ G {\bf 40}, 045001 (2013)


\bibitem{lhcb2} 
  R. Aaij {\it et al.}  [LHCb Collaboration],
   J.\ Phys.\ G {\bf 41}, 055002 (2014)
   
\bibitem{lhcb3} 
  R.~Aaij {\it et al.} [LHCb Collaboration],
  JHEP {\bf 1509}, 084 (2015)
 
\bibitem{lhcbconf} 
  R.~Aaij {\it et al.} [LHCb Collaboration],
  LHCb-CONF-2016-007.


\bibitem{review_forward} 
  K.~Akiba {\it et al.} [LHC Forward Physics Working Group Collaboration],
  J.\ Phys.\ G {\bf 43}, 110201 (2016)

\bibitem{bert}
  V.~P.~Goncalves and C.~A.~Bertulani,
  Phys.\ Rev.\ C {\bf 65}, 054905 (2002).
  

 \bibitem{vicmag}
  V.~P.~Goncalves and M.~V.~T.~Machado,
  Eur.\ Phys.\ J.\  C {\bf 40}, 519 (2005);   Phys.\ Rev.\  C {\bf 73}, 044902 (2006);
  Phys.\ Rev.\  D {\bf 77}, 014037 (2008);
  Phys.\ Rev.\ C {\bf 84}, 011902 (2011)


  \bibitem{brunoall} 
  V.~P.~Goncalves, B.~D.~Moreira and F.~S.~Navarra,
  Phys.\ Rev.\ C {\bf 90}, 015203 (2014);
 Phys.\ Lett.\ B {\bf 742}, 172 (2015).     

\bibitem{brunorun2} 
  V.~P.~Goncalves, B.~D.~Moreira and F.~S.~Navarra,
  Phys.\ Rev.\ D {\bf 95}, no. 5, 054011 (2017)  
  
\bibitem{vicdiego} 
  V.~P.~Goncalves, F.~S.~Navarra and D.~Spiering,
  Phys.\ Lett.\ B {\bf 768}, 299 (2017)


\bibitem{roman} 
  Y.~Hagiwara, Y.~Hatta, R.~Pasechnik, M.~Tasevsky and O.~Teryaev,
  Phys.\ Rev.\ D {\bf 96}, no. 3, 034009 (2017)



\bibitem{Dress} M.~Drees and D.~Zeppenfeld, Phys.\ Rev.\ D {\bf
39}, 2536 (1989).



\bibitem{Kniehl}
B.~A.~Kniehl,
Phys.\ Lett.\ B {\bf 254}, 267 (1991).



\bibitem{velhos_cem} 
  H.~Fritzsch,
  Phys.\ Lett.\  {\bf 67B}, 217 (1977); F.~Halzen,
  Phys.\ Lett.\  {\bf 69B}, 105 (1977).
 F.~Halzen and S.~Matsuda,
  Phys.\ Rev.\ D {\bf 17}, 1344 (1978)

\bibitem{cem_gregores}  
   J.~F.~Amundson, O.~J.~P.~Eboli, E.~M.~Gregores and F.~Halzen,
  Phys.\ Lett.\ B {\bf 390}, 323 (1997);  O.~J.~P.~Eboli, E.~M.~Gregores and F.~Halzen,
  Phys.\ Lett.\ B {\bf 451}, 241 (1999); 
  Phys.\ Rev.\ D {\bf 67}, 054002 (2003). 
  
 \bibitem{criscem} 
  M.~B.~Gay Ducati and C.~Brenner Mariotto,
  Phys.\ Lett.\ B {\bf 464}, 286 (1999) 
 
 \bibitem{crisvic_cem} 
  M.~B.~Gay Ducati, V.~P.~Goncalves and C.~Brenner Mariotto,
  Phys.\ Rev.\ D {\bf 65}, 037503 (2002) 

\bibitem{ramona} 
  Y.~Q.~Ma and R.~Vogt,
  Phys.\ Rev.\ D {\bf 94}, no. 11, 114029 (2016);  V.~Cheung and R.~Vogt,
  Phys.\ Rev.\ D {\bf 95}, no. 7, 074021 (2017); Phys.\ Rev.\ D {\bf 96}, no. 5, 054014 (2017).


\bibitem{anselmino} 
  M.~Anselmino, U.~D'Alesio and F.~Murgia,
  Phys.\ Rev.\ D {\bf 67}, 074010 (2003)

\bibitem{gluck_reya} 
  M.~Gluck and E.~Reya,
  Phys.\ Lett.\  {\bf 79B}, 453 (1978).



\bibitem{pisano} 
  U.~D'Alesio, F.~Murgia and C.~Pisano,
  JHEP {\bf 1509}, 119 (2015)
  
\bibitem{boervolg} 
  D.~Boer and W.~Vogelsang,
  Phys.\ Rev.\ D {\bf 69}, 094025 (2004)  
  
  
\bibitem{cteq6}
J.~Pumplin, D.~R.~Stump, J.~Huston, H.~L.~Lai, P.~M.~Nadolsky and W.~K.~Tung,
  JHEP {\bf 0207}, 012 (2002)
  
  
\bibitem{vicmairon} 
  V.~P.~Goncalves and M.~M.~Machado,
  Eur.\ Phys.\ J.\ A {\bf 50}, 72 (2014)

\bibitem{brunoluizvictor} 
  V.~P.~Goncalves, L.~S.~Martins and B.~D.~Moreira,
  Phys.\ Rev.\ D {\bf 96}, no. 7, 074029 (2017).
  
\bibitem{after} 
  S.~J.~Brodsky, F.~Fleuret, C.~Hadjidakis and J.~P.~Lansberg,
  Phys.\ Rept.\  {\bf 522}, 239 (2013); J.~P.~Lansberg {\it et al.},
  PoS QNP {\bf 2012}, 049 (2012);     
J.~P.~Lansberg {\it et al.},
  PoS DIS {\bf 2016}, 241 (2016).    


\end{thebibliography}
\end{document}